\documentclass[11pt,a4paper,preprintnumbers,notitlepage,nofootinbib]{revtex4-1}

\usepackage[T1]{fontenc} 
\usepackage{graphicx}
\usepackage{xspace}
\usepackage{amsmath,amssymb,mathtools}
\RequirePackage{flushend}
\RequirePackage[colorlinks,citecolor=blue,urlcolor=blue,linkcolor=blue]{hyperref}
\usepackage[utf8]{inputenc}
\usepackage{color}
\usepackage{natbib}
\usepackage{enumitem}

\def\nn{\nonumber}
\newcommand{\veff}{V_{\text{eff}}}
\newcommand{\vtree}{V^{(0)}}
\newcommand{\vone}{V^{(1)}}
\newcommand{\vtwo}{V^{(2)}}
\newcommand{\vphys}{v_\text{phys}}

\newcommand{\delt}{\delta^{(2)}}

\def\msbar{{\ensuremath{\overline{\rm MS}}}\xspace}
\def\lt{\tilde\lambda}
\def\lh{\hat\lambda}

\begin{document}

\preprint{OU-HET-1041}

\title{Two-loop corrections to the Higgs trilinear coupling in models with extended scalar sectors\\\textit{{\mdseries{\small Talk presented at the International Workshop on Future Linear Colliders (LCWS2019), Sendai, Japan,\hspace{3cm} 28 October-1 November, 2019. C19-10-28.}}}}

\author{Johannes~Braathen\footnote{Speaker.}} 
\email{braathen@het.phys.sci.osaka-u.ac.jp}
\affiliation{Department of Physics, Osaka University, Toyonaka, Osaka 560-0043, Japan}

\author{Shinya~Kanemura} 
\email{kanemu@het.phys.sci.osaka-u.ac.jp}
\affiliation{Department of Physics, Osaka University, Toyonaka, Osaka 560-0043, Japan}

\begin{abstract}

The Higgs trilinear coupling $\lambda_{hhh}$ is of great importance to understand the structure of the Higgs sector and allows searching for indirect signs of Beyond-the-Standard-Model (BSM) physics, even if new states are somehow hidden. In particular, in models with extended Higgs sectors, it is known that non-decouplings effects in BSM-scalar contributions at one loop can cause $\lambda_{hhh}$ to deviate significantly from its SM prediction, raising the question of what happens at two loops. We review here our calculation~\cite{Braathen:2019pxr,Braathen:2019zoh} of the leading two-loop corrections to $\lambda_{hhh}$ in an aligned scenario of a Two-Higgs-Doublet Model. We find their typical size to be 10-20\% of the one-loop corrections, meaning that they do not modify significantly the one-loop non-decoupling effects, but are not entirely negligible either.

\end{abstract}

\maketitle

\section{Introduction}

The discovery of the 125-GeV Higgs boson at the CERN LHC in 2012 has been a great success for particle physics, and has completed the particle spectrum of the Standard Model (SM). However, while it is now established that the Higgs sector is responsible for the electroweak symmetry breaking (EWSB), very little is known about the nature of the Higgs potential -- in fact, only its minimum and the curvature around the minimum are known. Moreover, although some new physics must exist -- possibly not too far from the electroweak (EW) scale -- to address shortcomings of the SM, there is so far no clear sign of it in experiments. A first possible explanation for this could be that new physics only exists beyond the scale currently accessible at colliders, and hence their effects would be (almost) impossible to find. Another interesting possibility would be that the new states are made somehow difficult to observe via some symmetry or mechanism. 

A prime example of the latter is \textit{alignment}~\cite{Gunion:2002zf}, which is defined for models of several Higgs doublets as the limit in which the total EW vacuum expectation value (VEV) is colinear in field space with one of the CP-even Higgs mass eigenstates. This limit is a natural consequence of decoupling, but more interestingly it can also occur even without decoupling. In aligned scenarios, the couplings of the state carrying the EW VEV are, at tree level, exactly equal to the SM-Higgs couplings, while the other scalars do not couple to the weak gauge bosons and are thus difficult to detect. Nevertheless, radiative corrections to properties of the discovered 125-GeV Higgs boson (in particular its couplings) can include effects from BSM particles that cause (potentially large) deviations \textit{at loop level} from SM predictions, even if the new particles themselves are not observed -- this was discussed in Refs.~\cite{Kanemura:2002vm,Kanemura:2004mg}.
In this way, Higgs couplings offer an important opportunity to distinguish aligned BSM scenarios from the SM (or scenarios where new physics is decoupled). 

Among these couplings, one of the most important ones is the Higgs trilinear coupling $\lambda_{hhh}$. First of all, $\lambda_{hhh}$ determines the shape of the Higgs potential, and thus its magnitude is crucial in deciding the strength of the EWSB. More precisely, it was found in Refs.~\cite{Grojean:2004xa,*Kanemura:2004ch} that in a Two-Higgs-Doublet Model (2HDM) $\lambda_{hhh}$ must deviate by at least 20\% from its SM value for the EW phase transition to be of strong first order -- which is itself necessary for the success of the scenario of electroweak baryogenesis~\cite{Kuzmin:1985mm,*Cohen:1993nk}. Furthermore, while most couplings of the Higgs boson are now known to a good level of accuracy, deviations of several hundred percent are still allowed for $\lambda_{hhh}$: indeed the current best limit is given by ATLAS as $-3.7<\lambda_{hhh}^\text{exp.}/\lambda_{hhh}^\text{SM}<11.5$~\cite{ATLAS:2019pbo} (at 95\% confidence level). This situation will be improved at future colliders: it is expected that the HL-LHC will be able to constrain $\lambda_{hhh}$ to about 50\%~\cite{Cepeda:2019klc}, while lepton colliders (ILC, CLIC, etc.) could achieve precisions of some tens of percent~\cite{Fujii:2015jha,*Roloff:2019crr}. For more details, the reader may find a complete review in Ref.~\cite{deBlas:2019rxi}. 

One-loop expressions are now available for Higgs couplings (and decays) in a wide range of BSM models~\cite{Barger:1991ed,Hollik:2001px,Dobado:2002jz,Kanemura:2002vm,Kanemura:2004mg,Aoki:2012jj,Kanemura:2015mxa,Arhrib:2015hoa,Kanemura:2015fra,Krause:2016oke,Kanemura:2016sos,He:2016sqr,Kanemura:2016lkz,Kanemura:2017wtm}, and are implemented in public codes such as \texttt{H-COUP}~\cite{Kanemura:2017gbi,*Kanemura:2019slf} or \texttt{2HDECAY}~\cite{Krause:2018wmo}. Non-decoupling effects are common in these one-loop calculations, thereby posing the question of whether new large corrections can occur at two loops. For this reason, in Refs.~\cite{Braathen:2019pxr,Braathen:2019zoh}, we derived\footnote{The other existing works on two-loop contributions to $\lambda_{hhh}$ are computations of the leading two-loop SUSY-QCD corrections in the MSSM~\cite{Brucherseifer:2013qva} and the NMSSM~\cite{Muhlleitner:2015dua}, and the calculation of (part of) the dominant two-loop scalar effects in the Inert Doublet Model~\cite{Senaha:2018xek}. } the leading two-loop corrections to $\lambda_{hhh}$ in an aligned scenario of a 2HDM, in the Inert Doublet Model and in a singlet extension of the SM. We review our calculation in these proceedings, focusing on the case of the 2HDM. After recalling our calculational setup, we show numerical examples to evaluate the maximal magnitude that the two-loop corrections to $\lambda_{hhh}$ can reach. As we will see, while the new two-loop contributions remain well under control (at least while perturbative unitarity is maintained) and do not significantly alter the non-decoupling effects found at one loop, they turn out not to be entirely negligible either. 

\section{The Two-Higgs-Doublet Model}
We consider here a CP-conserving 2HDM, defined in terms of two $SU(2)_L$ doublets of hypercharge $1/2$ -- $\Phi_i=\begin{psmallmatrix} \Phi_i^+\\\Phi_i^0\end{psmallmatrix}$, $i=1,2$. We also introduce a softly-broken $\mathbb{Z}_2$ symmetry under which $\Phi_2$ flips sign, to avoid tree-level flavour-changing neutral currents. The tree-level potential is then 
\begin{align}
\vtree=&\ m_{1}^2 |\Phi_1|^2 + m_{2}^2 |\Phi_2|^2 -  m_3^2\big( \Phi_2^\dagger\Phi_1 + \Phi_1^\dagger\Phi_2\big) +\frac{\lambda_1}{2}|\Phi_1|^4 +\frac{\lambda_2}{2}|\Phi_2|^4 \nn\\ 
        &+ \lambda_3 |\Phi_1|^2|\Phi_2|^2 +\lambda_4 |\Phi_2^\dagger\Phi_1|^2  +\frac{\lambda_5}{2} \Big[ (\Phi_2^\dagger\Phi_1)^2+ (\Phi_1^\dagger\Phi_2)^2 \Big]\,.
\end{align}
where all parameters, as well as the scalar VEVs $v_i\equiv\langle\Phi_i^0\rangle/\sqrt{2}$, are real because the assumption of CP conservion. Mass eigenstates can be obtained (at tree level) from the components of $\Phi_{1,2}$ using rotations of angles $\alpha$ for the CP-even states and $\beta$ for the CP-odd and the charged states -- the latter being defined by the ratio $\tan\beta\equiv v_2/v_1$. There are then four physical Higgs states: two CP-even scalars $h$ and $H$, a CP-even pseudoscalar $A$, and a charged Higgs boson $H^\pm$. Among these, we consider the lightest CP-even mass eigenstate $h$ to be the discovered 125-GeV Higgs boson, while $H$, $A$, and $H^+$ are the additional BSM scalar 2HDM.

Of the ten parameters in the 2HDM scalar sector -- $m_1^2,\,m_2^2,\,m_3^2,\,\lambda_1,\,\lambda_2,\,\lambda_3,\,\lambda_4,\,\lambda_5,\,v_1,\,v_2$ -- two (typically $m_1^2,\,m_2^2$) are eliminated using the tadpole equations. Moreover, only the ratio of the scalar VEVs is undetermined (because the total EW VEV $v=\sqrt{v_1^2+v_2^2}\approx 246\text{ GeV}$ is known), and one of the scalar quartic couplings must be fixed in order to reproduce $m_h\simeq 125\text{ GeV}$. It is also common to trade Lagrangian parameters for other inputs. First, the five scalar quartic couplings are replaced\footnote{This conversion is performed at tree level, using relations that can be found for example in Ref.~\cite{Kanemura:2004mg}. It is then important to keep in mind that the scalar masses should then be considered as \textit{tree-level}, rather than pole, masses. The relation between Lagrangian scalar couplings and pole masses in the 2HDM is discussed $e.g.$ in Refs.~\cite{Kanemura:2015mxa,Krause:2016oke,Basler:2016obg,Kanemura:2017wtm,Braathen:2017izn,Braathen:2017jvs,Basler:2017uxn,Kanemura:2019kjg}. } by the four mass eigenvalues -- $m_h$, $m_H$, $m_A$, $m_{H^\pm}$ -- and the CP-even mixing angle $\alpha$. Next, the off-diagonal mass term $m_3^2$ is used to define a soft-breaking scale $M$ of the $\mathbb{Z}_2$-symmetry, as $M^2\equiv m_3^2/(c_\beta s_\beta)$. The remaining free parameters in the scalar sector are then
\begin{equation}
 M,\,m_H,\,m_A,\,m_{H^\pm},\,\alpha,\text{ and } \beta\ \text{(or }\tan\beta\text{)}\,.
\end{equation}

In the following, we will further consider that we work in the alignment limit~\cite{Gunion:2002zf}, as this considerably relaxes the experimental constraints on 2HDM scenarios. In field space, this is the limit in which the rotation of angle $\beta$ suffices to diagonalise not only the CP-odd and charged Higgs mass matrices but also the CP-even one. There are two possible choices then, either $\alpha=\beta$ or $\alpha=\beta-\pi/2$, of which we choose the latter so that the lightest CP-even state $h$ is the SM-like Higgs boson. 
It is straightforward to show that in the alignment limit, the field-dependent (tree-level) masses of the top quark and of each of the BSM scalars $\Phi=H,A,H^\pm$ can be written as
\begin{equation}
 m_t^2(h)=y_t^2s_\beta^2(v+h)^2/2,\quad\text{and}\quad m_\Phi^2(h)=M^2+\tilde\lambda_\Phi (v+h)^2\,,
\end{equation}
where the precise expressions for $\lt_\Phi$ are given $e.g.$ in eqs.~(II.9) and (II.13) of Ref.~\cite{Braathen:2019zoh}.

\section{One-loop corrections to $\lambda_{hhh}$ and non-decoupling effects}
The expression of $\lambda_{hhh}$ to leading one-loop order, and in the alignment limit, reads~\cite{Kanemura:2002vm,Kanemura:2004mg}
\begin{equation}
 \lambda_{hhh}=\frac{3[M_h^2]_{\veff}}{v}+\frac{1}{16\pi^2}\Bigg[-\frac{48m_t^4}{v^3}+\sum_{\Phi=H,A,H^\pm}\frac{4n_\Phi m_\Phi^4}{v^3}\left(1-\frac{M^2}{m_\Phi^2}\right)^3\Bigg]+\cdots\,,
\end{equation}
where $[M_h^2]_{\veff}$ denotes the Higgs effective-potential mass (we return to this point below). 
The one-loop part therein contains two types of terms: the first corresponding to the SM-like top quark loop, and the second to loops of BSM scalars ($H$, $A$, $H^\pm$). The behaviour of the latter is crucially determined by the reduction factor $(1-M^2/m_\Phi^2)$. Indeed, we have that
\begin{equation}
\label{EQ:red_fact}
 \left(1-\frac{M^2}{m_\Phi^2}\right)=\frac{\lt_\Phi v^2}{M^2+\lt_\Phi v^2}\to\left\{\begin{matrix*}[l] 0 \text{ if  }M^2\gg \tilde\lambda_\Phi v^2\text{, case \textit{(i)}}\\ 1 \text{ if }M^2\ll \tilde\lambda_\Phi v^2\text{, case \textit{(ii)}}\end{matrix*}\right.\,,
\end{equation}
so that if the parameter $M$ is small, or in other words if the BSM scalars acquire their masses mostly from the EWSB, the BSM contributions to $\lambda_{hhh}$ scale like $m_\Phi^4$. In this case, the BSM corrections can cause the Higgs trilinear coupling in the 2HDM to deviate by several tens or even more than a hundred percent from its SM prediction, and so even well before violating perturbative unitarity \cite{Lee:1977eg,Kanemura:1993hm,Akeroyd:2000wc}. Such non-decoupling effects\footnote{These effects appear when the BSM scalars acquire their masses via the EWSB and thus cannot be decoupled from the theory, hence the name \textit{non-decoupling effects}. } were first found for the 2HDM in Ref.~\cite{Kanemura:2002vm}, and are now known to be present in Higgs-boson couplings in various non-supersymmetric extensions of the SM~\cite{Kanemura:2002vm,Kanemura:2004mg,Aoki:2012jj,Krause:2016oke,Kanemura:2015mxa,Arhrib:2015hoa,Kanemura:2015fra,Kanemura:2016sos,He:2016sqr,Kanemura:2016lkz,Kanemura:2017wtm}.

It is important to note furthermore that having one-loop corrections of the same order as the tree-level value of $\lambda_{hhh}$ does not signal a problem in the perturbative expansion. Indeed, the loop corrections involve a new parameter -- namely $m_\Phi$ -- that was not present at the tree level (which only involves $m_h$ in the alignment limit), and are therefore not a perturbation of the tree-level expression. Nevertheless, one may legitimately ask what can then happen at two loops, and whether new large effects can once again appear. This is what we will consider in the next section.

\section{Two-loop corrections to $\lambda_{hhh}$ and numerical analysis}

To derive the leading two-loop corrections to $\lambda_{hhh}$, we follow the following three steps:
\begin{itemize}[itemsep=-2pt,topsep=1pt]
 \item[(i)]   We calculate the effective potential $\veff$ up to leading corrections at two loops, $i.e.$
 \begin{equation}
  \veff=\vtree+\frac{1}{16\pi^2}\vone+\frac{1}{(16\pi^2)^2}\vtwo+\cdots\,.
 \end{equation}
 The two-loop contributions $\vtwo$ are given by one-particle-irreducible vacuum bubble diagrams, which can be computed -- in the Landau gauge and in terms of \msbar -renormalised parameters -- using results from Refs.~\cite{Martin:2001vx,*Martin:2003qz,*Degrassi:2009yq,*Braathen:2016cqe}. 
 \item[(ii)]  From the effective potential, we can compute an effective Higgs trilinear coupling $\lambda_{hhh}$ as
 \begin{equation}
 \label{EQ:def_eff_coup}
  \lambda_{hhh}\equiv\frac{\partial^3\veff}{\partial h^3}\bigg|_\text{min.}=\frac{3[M_h^2]_{\veff}}{v}+\left[\frac{3}{v}\left(\frac{1}{v}\frac{\partial}{\partial h}-\frac{\partial^2}{\partial h^2}\right)+\frac{\partial^3}{\partial h^3}\right]\veff\bigg|_\text{min.}\,.
 \end{equation}
 This effective coupling differs from the full Higgs three-point function in that it does not include the sub-leading dependence on external momenta. The second equality in eq.~(\ref{EQ:def_eff_coup}) serves to replace the tree-level mass of the lightest Higgs boson by its effective-potential mass and also ensures that the minimisation conditions of the full potential are taken into account. 
 \item[(iii)] The last step in our calculation is to express our results in terms of on-shell (OS) quantities that can be related to experimental data, namely the pole masses of the BSM scalars $M_H,\,M_A,\,M_{H^\pm}$ and of the top quark $M_t$, as well as the physical EW VEV $\vphys=(\sqrt{2}G_F)^{-1/2}$ ($G_F$ being the Fermi constant). Including also the finite effects from wave-function renormalisation (WFR), we obtain our on-shell scheme result for the Higgs trilinear coupling, which we denote with a hat $\lh_{hhh}$ to distinguish it from its \msbar counterpart, as
 \begin{equation}
  \hat\lambda_{hhh}\equiv \left(\frac{1+\delta Z_h^\text{OS}}{1+\delta Z_h^\msbar}\right)^{3/2}\lambda_{hhh}\,,
 \end{equation}
 where $\delta Z_h^{\text{OS}/\msbar}$ are the OS and \msbar WFR counterterms for the Higgs field, and where $\lambda_{hhh}$ is expressed in terms of OS-renormalised quantities. All the necessary shifts to replace \msbar parameters by OS ones can be found $e.g.$ in Ref.~\cite{Braathen:2019zoh}. 
\end{itemize}

The main purpose of our calculation is to estimate the maximal possible size of the two-loop corrections and, for this reason, we have the freedom to neglect subleading effects here. Therefore, in addition to the external-momentum dependence mentioned above, we also neglect contributions from the lighter scalars -- that is the 125-GeV Higgs boson and the would-be Goldstone bosons -- as well as from light fermions, and consider only effects from the (heavy) BSM scalars and from the top quark. In addition to this, we do not include loop-induced deviations from the alignment condition $s_{\beta-\alpha}=1$ (such effects were found in Ref.~\cite{Braathen:2017izn} to be moderate).  

Following the above steps, we obtain the for the leading two-loop corrections to $\lh_{hhh}$ in terms of on-shell-renormalised quantities
\begin{align}
\label{EQ:2HDM_OStres}
 \delt\hat\lambda_{hhh}=\frac{3 M_\Phi^6}{16\pi^4\vphys^5}\Bigg\{&\left(1-\frac{\tilde M^2}{M_\Phi^2}\right)^4 \left\{4+3\cot^22\beta\left[3-\frac{\pi}{\sqrt{3}}\left(\frac{\tilde M^2}{M_\Phi^2} + 2\right)\right]\right\}\nn\\
 &+12\cot^22\beta \left(1-\frac{\tilde M^2}{M_\Phi^2}\right)^4+\frac{6 M_t^2\cot^2\beta}{M_\Phi^2}\left(1-\frac{\tilde M^2}{M_\Phi^2}\right)^3\nn\\
 &-\left(1-\frac{\tilde M^2}{M_\Phi^2}\right)^5+\frac{7M_t^2}{2M_\Phi^2}\left(1-\frac{\tilde M^2}{M_\Phi^2}\right)^3\Bigg\}+\mathcal{O}\left(\frac{M_\Phi^2M_t^4}{256\pi^4\vphys^5}\right)\,,
\end{align}
where for brievety we have taken the pole masses of the BSM scalars to be degenerate ($M_H=M_A=M_{H^\pm}=M_\Phi$), and we have not written out explicitly terms of order $M_\Phi^2M_t^4$ that become subdominant for $M_\Phi\gtrsim M_t$ (complete expressions can be found in Appendix B of Ref.~\cite{Braathen:2019zoh}). The parameter $\tilde M$ is defined from $M$ in such a way as to ensure the proper decoupling of the BSM corrections in the OS scheme when using a relation of the form $M_\Phi^2=\tilde M^2+\lt_\Phi v^2$ -- for details on our prescription for $\tilde M$, we refer the reader to the discussions in Refs.~\cite{Braathen:2019pxr,Braathen:2019zoh}. 

\begin{figure}[t]
 \centering
 \includegraphics[width=.495\textwidth]{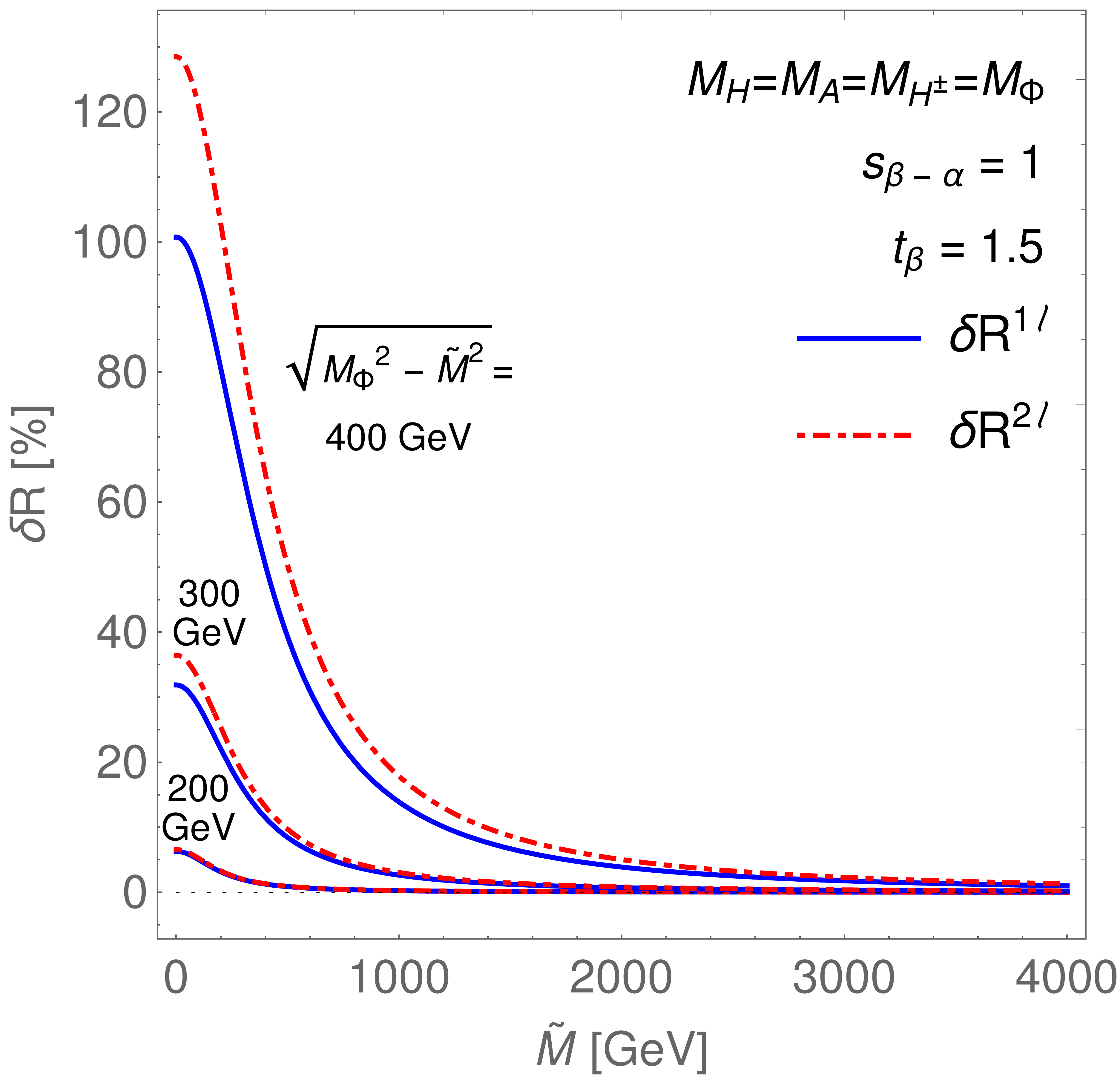}
 \includegraphics[width=.49\textwidth]{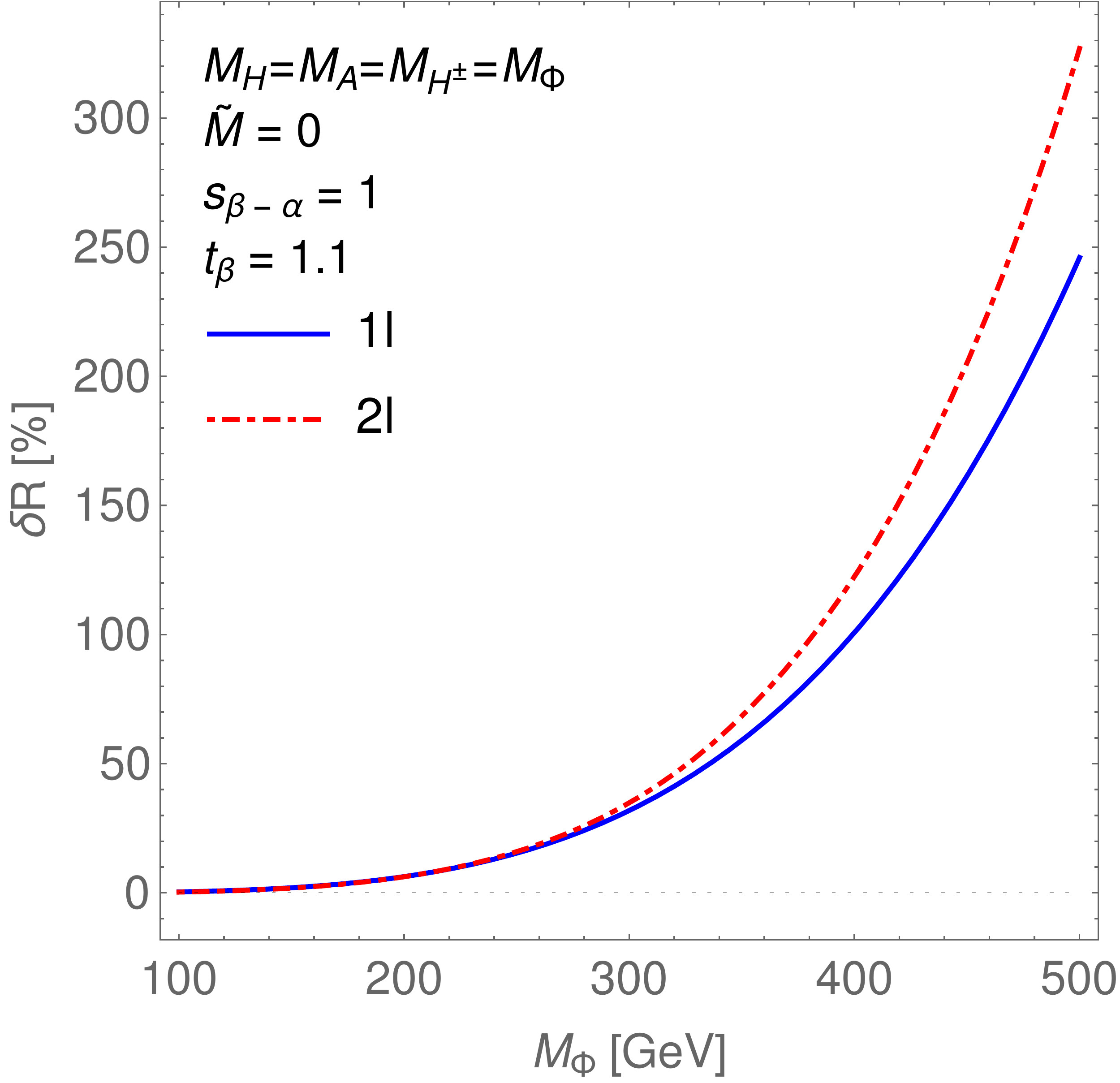}
 \caption{Illustrations of the behaviour of the BSM deviation $\delta R\equiv (\lh_{hhh}^\text{2HDM}-\lh_{hhh}^\text{SM})/\lh_{hhh}^\text{SM}$ at one loop (\textit{solid blue curves}) and at two loops (\textit{dot-dashed red curves}). (\textbf{Left:}) Results for $\delta R$ as a function of $\tilde M$, for the three values of $\lt_\Phi v^2=M_\Phi^2-\tilde M^2=(200\text{ GeV})^2,\,(300\text{ GeV})^2,\,(400\text{ GeV})^2$. (\textbf{Right:}) $\delta R$ as a function of the degenerate mass of the 2HDM scalars $M_\Phi$.}
 \label{FIG:2HDM1}
\end{figure}

Turning now to the numerical study of our results, the first point we can investigate is the decoupling behaviour of the two-loop corrections. The left side of figure~\ref{FIG:2HDM1} shows the ratio $\delta R\equiv (\lh_{hhh}^\text{2HDM}-\lh_{hhh}^\text{SM})/\lh_{hhh}^\text{SM}$ as a function of $\tilde M$ at one loop (blue) and two loops (red), for different values of $\lt_\Phi v^2$ and $\tan\beta=1.5$. The BSM deviations can be seen to tend to zero rapidly, thereby confirming that $\tilde M$ is the proper parameter to control the decoupling of the BSM loop corrections, when these are expressed in terms of OS quantities.  

The other limit of interest is when the BSM effects are maximal, $i.e.$ for $\tilde M=0$. The right side of figure~\ref{FIG:2HDM1} illustrates this case, showing once again the ratio $\delta R$, at both one- and two-loop orders, as a function of $M_\Phi$, for $\tilde M=0$ and $\tan\beta=1.1$. At first one may observe that the two-loop corrections grow faster than their one-loop counterparts -- indeed the dominant terms at two loops scale like $M_\Phi^6$ rather than like $M_\Phi^4$ as at one loop. However, it is important to note that the size of the two-loop corrections remains well below that of the one-loop ones, for the whole range of BSM scalar masses considered here where we have verified that (tree-level) perturbative unitarity~\cite{Lee:1977eg,Kanemura:1993hm,Akeroyd:2000wc} is satisfied. If we consider the size of the BSM deviations for $M_\Phi=300\text{ GeV}$ and $400\text{ GeV}$ ($i.e.$ well below the bound from perturbative unitarity), we find that two-loop corrections are then respectively about 10\% and 20\% of their one-loop counterparts. We keep these as typical values, as for larger $M_\Phi$ we start approaching the unitarity bound (and then our calculation might become less reliable).

\begin{figure}[ht]
 \centering
 \hspace{-1cm}\includegraphics[width=.6\textwidth]{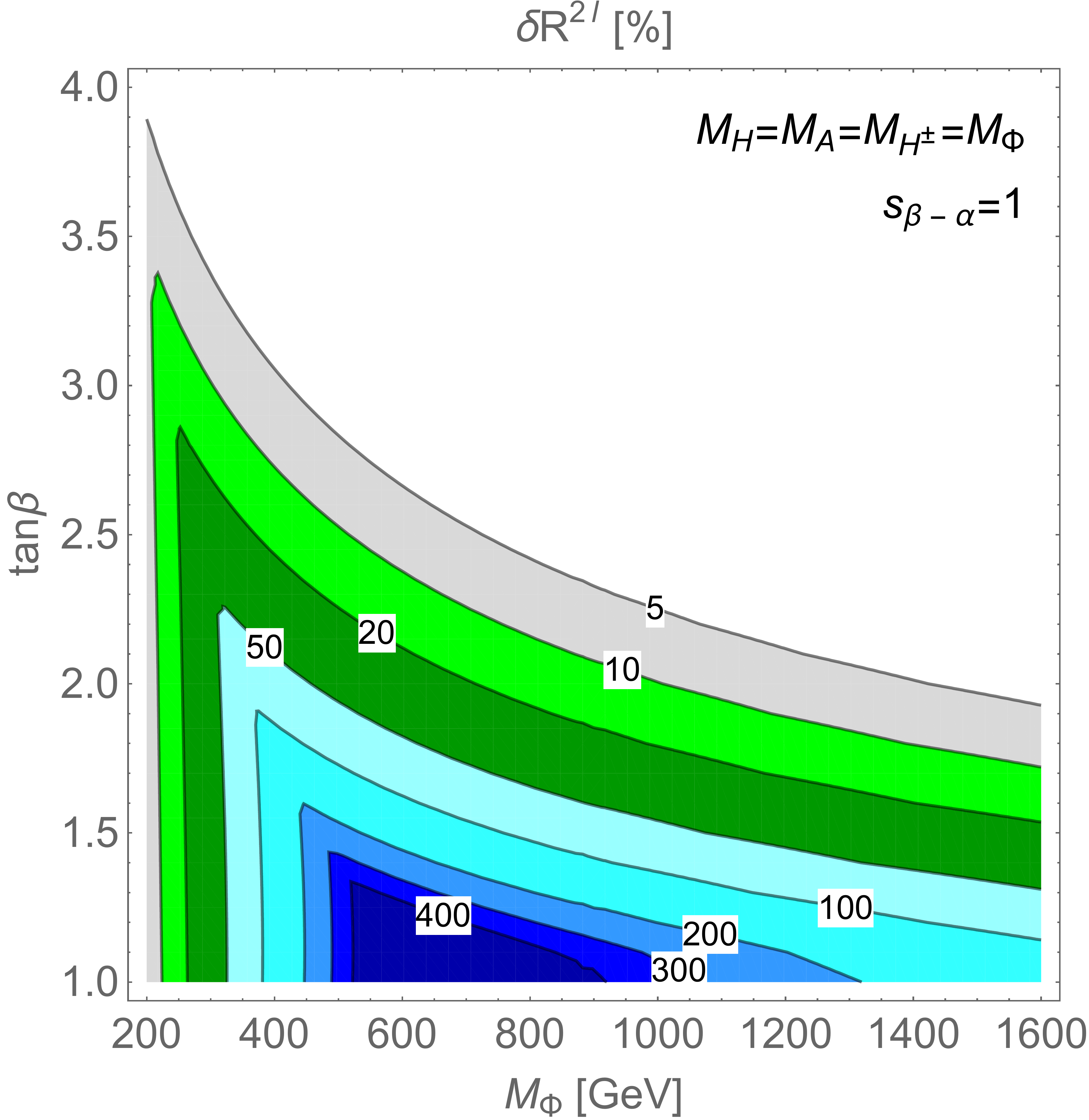}
 \caption{Maximal deviation $\delta R$ of the Higgs trilinear coupling $\lh_{hhh}$ in the 2HDM from its SM prediction allowed under the criterion of tree-level perturbative unitarity, in the plane of $M_\Phi$ and $\tan\beta$.}
 \label{FIG:contour}
\end{figure}

Lastly, the contour plot in figure~\ref{FIG:contour} shows in the plane of $\tan\beta$ and $M_\Phi$ the maximal BSM deviation $\delta R$ that can be obtained at two loops while fulfilling the criterion of (tree-level) perturbative unitarity. For a given point in the $\{\tan\beta,M_\Phi\}$ plane, the only parameter left to decide how large the corrections can become is $\tilde M$, which should be as small as possible. The perturbative unitarity conditions essentially set upper bounds on the allowed values of the quartic coupling, and hence indirectly constrain how small $\tilde M$ can be (as $M_\Phi^2=\tilde M^2+\lt_\Phi v^2$ is fixed). For lower values of $M_\Phi$ -- $e.g.$ $M_\Phi\lesssim 700\text{ GeV }(600\text{ GeV})$ for $\tan\beta=1\ (1.2)$ -- the bound on the quartic coupling is not reached and $\tilde M=0$ is allowed, so that $\delta R$ grows with $M_\Phi$ as seen also in the right plot of fig.~\ref{FIG:2HDM1}. For larger values of $M_\Phi$, $\tilde M$ cannot remain zero and must take a finite value, which in turn means that the BSM deviations decrease due to the reduction factors ($c.f.$ eq.~(\ref{EQ:red_fact})) and eventually vanish in the large-mass limit (as required by the decoupling theorem). Note also that the value of $M_\Phi$ until which $\tilde M$ can remain zero decreases with $\tan\beta$ because the tree-level unitarity conditions become more stringent for increasing $\tan\beta$. In consequence, the largest BSM deviations are found for intermediate values of the BSM scalar masses and for low $\tan\beta$. Finally, concerning the reach of future experiments, the HL-LHC should be able to access (roughly) the regions shaded in blue~\cite{Cepeda:2019klc}, while future lepton colliders -- $e.g.$ ILC or CLIC -- could probe the green-shaded regions~\cite{Fujii:2015jha,*Roloff:2019crr}.  

\section{Conclusion}
We have summarised here the results of the first calculation of two-loop corrections to the Higgs trilinear coupling in an aligned scenario of a 2HDM, performed in Refs.~\cite{Braathen:2019pxr,Braathen:2019zoh}. We find that the two-loop corrections remain smaller than their one-loop counterparts, at least as long as perturbative unitarity is verified. The typical size we find for the new contributions at two loops is of the order of 10-20\% of those at one loop. On the one hand, this confirms that the non-decoupling effects found at one loop are not drastically altered when including two-loop corrections. But, on the other hand, it also means that in the perspective of future precise measurements of the Higgs trilinear coupling, theory predictions beyond one loop will be needed. We further discussed here the regions of parameter space where the BSM deviation of $\lambda_{hhh}$ from its SM prediction are the largest, and what the reach of future experiments will be. This study should also serve as an example of how precise computations of Higgs-boson properties allow distinguishing aligned scenarios from decoupled ones. Finally, many interesting directions remain to continue this work, in particular tackling technical challenges and investigating corrections to other couplings of the Higgs boson.

\vspace{-.5cm}
\subsection*{Acknowledgments}\vspace{-.5cm}
This work is, in part, supported by Grant-in-Aid for Scientific Research on Innovative Areas, the Ministry of Education, Culture, Sports, Science and Technology, No. 16H06492 and No. 18H04587. This work is also supported in part by JSPS KAKENHI Grant No. A18F180220. \vspace{-.5cm}

\bibliography{BKproc}

\begin{thebibliography}{44}%
\makeatletter
\providecommand \@ifxundefined [1]{%
 \@ifx{#1\undefined}
}%
\providecommand \@ifnum [1]{%
 \ifnum #1\expandafter \@firstoftwo
 \else \expandafter \@secondoftwo
 \fi
}%
\providecommand \@ifx [1]{%
 \ifx #1\expandafter \@firstoftwo
 \else \expandafter \@secondoftwo
 \fi
}%
\providecommand \natexlab [1]{#1}%
\providecommand \enquote  [1]{``#1''}%
\providecommand \bibnamefont  [1]{#1}%
\providecommand \bibfnamefont [1]{#1}%
\providecommand \citenamefont [1]{#1}%
\providecommand \href@noop [0]{\@secondoftwo}%
\providecommand \href [0]{\begingroup \@sanitize@url \@href}%
\providecommand \@href[1]{\@@startlink{#1}\@@href}%
\providecommand \@@href[1]{\endgroup#1\@@endlink}%
\providecommand \@sanitize@url [0]{\catcode `\\12\catcode `\$12\catcode
  `\&12\catcode `\#12\catcode `\^12\catcode `\_12\catcode `\%12\relax}%
\providecommand \@@startlink[1]{}%
\providecommand \@@endlink[0]{}%
\providecommand \url  [0]{\begingroup\@sanitize@url \@url }%
\providecommand \@url [1]{\endgroup\@href {#1}{\urlprefix }}%
\providecommand \urlprefix  [0]{URL }%
\providecommand \Eprint [0]{\href }%
\providecommand \doibase [0]{http://dx.doi.org/}%
\providecommand \selectlanguage [0]{\@gobble}%
\providecommand \bibinfo  [0]{\@secondoftwo}%
\providecommand \bibfield  [0]{\@secondoftwo}%
\providecommand \translation [1]{[#1]}%
\providecommand \BibitemOpen [0]{}%
\providecommand \bibitemStop [0]{}%
\providecommand \bibitemNoStop [0]{.\EOS\space}%
\providecommand \EOS [0]{\spacefactor3000\relax}%
\providecommand \BibitemShut  [1]{\csname bibitem#1\endcsname}%
\let\auto@bib@innerbib\@empty
\bibitem [{\citenamefont {Braathen}\ and\ \citenamefont
  {Kanemura}(2019{\natexlab{a}})}]{Braathen:2019pxr}%
  \BibitemOpen
  \bibfield  {author} {\bibinfo {author} {\bibfnamefont {J.}~\bibnamefont
  {Braathen}}\ and\ \bibinfo {author} {\bibfnamefont {S.}~\bibnamefont
  {Kanemura}},\ }\href {\doibase 10.1016/j.physletb.2019.07.021} {\bibfield
  {journal} {\bibinfo  {journal} {Phys. Lett.}\ }\textbf {\bibinfo {volume}
  {B796}},\ \bibinfo {pages} {38} (\bibinfo {year} {2019}{\natexlab{a}})},\
  \Eprint {http://arxiv.org/abs/1903.05417} {arXiv:1903.05417 [hep-ph]}
  \BibitemShut {NoStop}%
\bibitem [{\citenamefont {Braathen}\ and\ \citenamefont
  {Kanemura}(2019{\natexlab{b}})}]{Braathen:2019zoh}%
  \BibitemOpen
  \bibfield  {author} {\bibinfo {author} {\bibfnamefont {J.}~\bibnamefont
  {Braathen}}\ and\ \bibinfo {author} {\bibfnamefont {S.}~\bibnamefont
  {Kanemura}},\ }\href@noop {} {\  (\bibinfo {year} {2019}{\natexlab{b}})},\
  \Eprint {http://arxiv.org/abs/1911.11507} {arXiv:1911.11507 [hep-ph]}
  \BibitemShut {NoStop}%
\bibitem [{\citenamefont {Gunion}\ and\ \citenamefont
  {Haber}(2003)}]{Gunion:2002zf}%
  \BibitemOpen
  \bibfield  {author} {\bibinfo {author} {\bibfnamefont {J.~F.}\ \bibnamefont
  {Gunion}}\ and\ \bibinfo {author} {\bibfnamefont {H.~E.}\ \bibnamefont
  {Haber}},\ }\href {\doibase 10.1103/PhysRevD.67.075019} {\bibfield  {journal}
  {\bibinfo  {journal} {Phys. Rev.}\ }\textbf {\bibinfo {volume} {D67}},\
  \bibinfo {pages} {075019} (\bibinfo {year} {2003})},\ \Eprint
  {http://arxiv.org/abs/hep-ph/0207010} {arXiv:hep-ph/0207010 [hep-ph]}
  \BibitemShut {NoStop}%
\bibitem [{\citenamefont {Kanemura}\ \emph {et~al.}(2003)\citenamefont
  {Kanemura}, \citenamefont {Kiyoura}, \citenamefont {Okada}, \citenamefont
  {Senaha},\ and\ \citenamefont {Yuan}}]{Kanemura:2002vm}%
  \BibitemOpen
  \bibfield  {author} {\bibinfo {author} {\bibfnamefont {S.}~\bibnamefont
  {Kanemura}}, \bibinfo {author} {\bibfnamefont {S.}~\bibnamefont {Kiyoura}},
  \bibinfo {author} {\bibfnamefont {Y.}~\bibnamefont {Okada}}, \bibinfo
  {author} {\bibfnamefont {E.}~\bibnamefont {Senaha}}, \ and\ \bibinfo {author}
  {\bibfnamefont {C.~P.}\ \bibnamefont {Yuan}},\ }\href {\doibase
  10.1016/S0370-2693(03)00268-5} {\bibfield  {journal} {\bibinfo  {journal}
  {Phys. Lett.}\ }\textbf {\bibinfo {volume} {B558}},\ \bibinfo {pages} {157}
  (\bibinfo {year} {2003})},\ \Eprint {http://arxiv.org/abs/hep-ph/0211308}
  {arXiv:hep-ph/0211308} \BibitemShut {NoStop}%
\bibitem [{\citenamefont {Kanemura}\ \emph {et~al.}(2004)\citenamefont
  {Kanemura}, \citenamefont {Okada}, \citenamefont {Senaha},\ and\
  \citenamefont {Yuan}}]{Kanemura:2004mg}%
  \BibitemOpen
  \bibfield  {author} {\bibinfo {author} {\bibfnamefont {S.}~\bibnamefont
  {Kanemura}}, \bibinfo {author} {\bibfnamefont {Y.}~\bibnamefont {Okada}},
  \bibinfo {author} {\bibfnamefont {E.}~\bibnamefont {Senaha}}, \ and\ \bibinfo
  {author} {\bibfnamefont {C.~P.}\ \bibnamefont {Yuan}},\ }\href {\doibase
  10.1103/PhysRevD.70.115002} {\bibfield  {journal} {\bibinfo  {journal} {Phys.
  Rev.}\ }\textbf {\bibinfo {volume} {D70}},\ \bibinfo {pages} {115002}
  (\bibinfo {year} {2004})},\ \Eprint {http://arxiv.org/abs/hep-ph/0408364}
  {arXiv:hep-ph/0408364} \BibitemShut {NoStop}%
\bibitem [{\citenamefont {Grojean}\ \emph {et~al.}(2005)\citenamefont
  {Grojean}, \citenamefont {Servant},\ and\ \citenamefont
  {Wells}}]{Grojean:2004xa}%
  \BibitemOpen
  \bibfield  {author} {\bibinfo {author} {\bibfnamefont {C.}~\bibnamefont
  {Grojean}}, \bibinfo {author} {\bibfnamefont {G.}~\bibnamefont {Servant}}, \
  and\ \bibinfo {author} {\bibfnamefont {J.~D.}\ \bibnamefont {Wells}},\ }\href
  {\doibase 10.1103/PhysRevD.71.036001} {\bibfield  {journal} {\bibinfo
  {journal} {Phys. Rev.}\ }\textbf {\bibinfo {volume} {D71}},\ \bibinfo {pages}
  {036001} (\bibinfo {year} {2005})},\ \Eprint
  {http://arxiv.org/abs/hep-ph/0407019} {arXiv:hep-ph/0407019} \BibitemShut
  {NoStop}%
\bibitem [{\citenamefont {Kanemura}\ \emph {et~al.}(2005)\citenamefont
  {Kanemura}, \citenamefont {Okada},\ and\ \citenamefont
  {Senaha}}]{Kanemura:2004ch}%
  \BibitemOpen
  \bibfield  {author} {\bibinfo {author} {\bibfnamefont {S.}~\bibnamefont
  {Kanemura}}, \bibinfo {author} {\bibfnamefont {Y.}~\bibnamefont {Okada}}, \
  and\ \bibinfo {author} {\bibfnamefont {E.}~\bibnamefont {Senaha}},\ }\href
  {\doibase 10.1016/j.physletb.2004.12.004} {\bibfield  {journal} {\bibinfo
  {journal} {Phys. Lett.}\ }\textbf {\bibinfo {volume} {B606}},\ \bibinfo
  {pages} {361} (\bibinfo {year} {2005})},\ \Eprint
  {http://arxiv.org/abs/hep-ph/0411354} {arXiv:hep-ph/0411354} \BibitemShut
  {NoStop}%
\bibitem [{\citenamefont {Kuzmin}\ \emph {et~al.}(1985)\citenamefont {Kuzmin},
  \citenamefont {Rubakov},\ and\ \citenamefont {Shaposhnikov}}]{Kuzmin:1985mm}%
  \BibitemOpen
  \bibfield  {author} {\bibinfo {author} {\bibfnamefont {V.~A.}\ \bibnamefont
  {Kuzmin}}, \bibinfo {author} {\bibfnamefont {V.~A.}\ \bibnamefont {Rubakov}},
  \ and\ \bibinfo {author} {\bibfnamefont {M.~E.}\ \bibnamefont
  {Shaposhnikov}},\ }\href {\doibase 10.1016/0370-2693(85)91028-7} {\bibfield
  {journal} {\bibinfo  {journal} {Phys. Lett.}\ }\textbf {\bibinfo {volume}
  {155B}},\ \bibinfo {pages} {36} (\bibinfo {year} {1985})}\BibitemShut
  {NoStop}%
\bibitem [{\citenamefont {Cohen}\ \emph {et~al.}(1993)\citenamefont {Cohen},
  \citenamefont {Kaplan},\ and\ \citenamefont {Nelson}}]{Cohen:1993nk}%
  \BibitemOpen
  \bibfield  {author} {\bibinfo {author} {\bibfnamefont {A.~G.}\ \bibnamefont
  {Cohen}}, \bibinfo {author} {\bibfnamefont {D.~B.}\ \bibnamefont {Kaplan}}, \
  and\ \bibinfo {author} {\bibfnamefont {A.~E.}\ \bibnamefont {Nelson}},\
  }\href {\doibase 10.1146/annurev.ns.43.120193.000331} {\bibfield  {journal}
  {\bibinfo  {journal} {Ann. Rev. Nucl. Part. Sci.}\ }\textbf {\bibinfo
  {volume} {43}},\ \bibinfo {pages} {27} (\bibinfo {year} {1993})},\ \Eprint
  {http://arxiv.org/abs/hep-ph/9302210} {arXiv:hep-ph/9302210} \BibitemShut
  {NoStop}%
\bibitem [{\citenamefont {collaboration}(2019)}]{ATLAS:2019pbo}%
  \BibitemOpen
  \bibfield  {author} {\bibinfo {author} {\bibfnamefont {T.~A.}\ \bibnamefont
  {collaboration}} (\bibinfo {collaboration} {ATLAS}),\ }\bibinfo
  {organization} {CERN}\ (\bibinfo  {publisher} {CERN},\ \bibinfo {address}
  {Geneva},\ \bibinfo {year} {2019})\ \bibinfo {note} {see also references
  herein.}\BibitemShut {Stop}%
\bibitem [{\citenamefont {Cepeda}\ \emph {et~al.}(2019)\citenamefont {Cepeda}
  \emph {et~al.}}]{Cepeda:2019klc}%
  \BibitemOpen
  \bibfield  {author} {\bibinfo {author} {\bibfnamefont {M.}~\bibnamefont
  {Cepeda}} \emph {et~al.} (\bibinfo {collaboration} {Physics of the HL-LHC
  Working Group}),\ }\href@noop {} {\  (\bibinfo {year} {2019})},\ \Eprint
  {http://arxiv.org/abs/1902.00134} {arXiv:1902.00134 [hep-ph]} \BibitemShut
  {NoStop}%
\bibitem [{\citenamefont {Fujii}\ \emph {et~al.}(2015)\citenamefont {Fujii}
  \emph {et~al.}}]{Fujii:2015jha}%
  \BibitemOpen
  \bibfield  {author} {\bibinfo {author} {\bibfnamefont {K.}~\bibnamefont
  {Fujii}} \emph {et~al.},\ }\href@noop {} {\  (\bibinfo {year} {2015})},\
  \Eprint {http://arxiv.org/abs/1506.05992} {arXiv:1506.05992 [hep-ex]}
  \BibitemShut {NoStop}%
\bibitem [{\citenamefont {Roloff}\ \emph {et~al.}(2019)\citenamefont {Roloff},
  \citenamefont {Schnoor}, \citenamefont {Simoniello},\ and\ \citenamefont
  {Xu}}]{Roloff:2019crr}%
  \BibitemOpen
  \bibfield  {author} {\bibinfo {author} {\bibfnamefont {P.}~\bibnamefont
  {Roloff}}, \bibinfo {author} {\bibfnamefont {U.}~\bibnamefont {Schnoor}},
  \bibinfo {author} {\bibfnamefont {R.}~\bibnamefont {Simoniello}}, \ and\
  \bibinfo {author} {\bibfnamefont {B.}~\bibnamefont {Xu}} (\bibinfo
  {collaboration} {CLICdp}),\ }\href@noop {} {\  (\bibinfo {year} {2019})},\
  \Eprint {http://arxiv.org/abs/1901.05897} {arXiv:1901.05897 [hep-ex]}
  \BibitemShut {NoStop}%
\bibitem [{\citenamefont {de~Blas}\ \emph {et~al.}(2019)\citenamefont {de~Blas}
  \emph {et~al.}}]{deBlas:2019rxi}%
  \BibitemOpen
  \bibfield  {author} {\bibinfo {author} {\bibfnamefont {J.}~\bibnamefont
  {de~Blas}} \emph {et~al.},\ }\href@noop {} {\  (\bibinfo {year} {2019})},\
  \Eprint {http://arxiv.org/abs/1905.03764} {arXiv:1905.03764 [hep-ph]}
  \BibitemShut {NoStop}%
\bibitem [{\citenamefont {Barger}\ \emph {et~al.}(1992)\citenamefont {Barger},
  \citenamefont {Berger}, \citenamefont {Stange},\ and\ \citenamefont
  {Phillips}}]{Barger:1991ed}%
  \BibitemOpen
  \bibfield  {author} {\bibinfo {author} {\bibfnamefont {V.~D.}\ \bibnamefont
  {Barger}}, \bibinfo {author} {\bibfnamefont {M.~S.}\ \bibnamefont {Berger}},
  \bibinfo {author} {\bibfnamefont {A.~L.}\ \bibnamefont {Stange}}, \ and\
  \bibinfo {author} {\bibfnamefont {R.~J.~N.}\ \bibnamefont {Phillips}},\
  }\href {\doibase 10.1103/PhysRevD.45.4128} {\bibfield  {journal} {\bibinfo
  {journal} {Phys. Rev.}\ }\textbf {\bibinfo {volume} {D45}},\ \bibinfo {pages}
  {4128} (\bibinfo {year} {1992})}\BibitemShut {NoStop}%
\bibitem [{\citenamefont {Hollik}\ and\ \citenamefont
  {Penaranda}(2002)}]{Hollik:2001px}%
  \BibitemOpen
  \bibfield  {author} {\bibinfo {author} {\bibfnamefont {W.}~\bibnamefont
  {Hollik}}\ and\ \bibinfo {author} {\bibfnamefont {S.}~\bibnamefont
  {Penaranda}},\ }\href {\doibase 10.1007/s100520100862} {\bibfield  {journal}
  {\bibinfo  {journal} {Eur. Phys. J.}\ }\textbf {\bibinfo {volume} {C23}},\
  \bibinfo {pages} {163} (\bibinfo {year} {2002})},\ \Eprint
  {http://arxiv.org/abs/hep-ph/0108245} {arXiv:hep-ph/0108245} \BibitemShut
  {NoStop}%
\bibitem [{\citenamefont {Dobado}\ \emph {et~al.}(2002)\citenamefont {Dobado},
  \citenamefont {Herrero}, \citenamefont {Hollik},\ and\ \citenamefont
  {Penaranda}}]{Dobado:2002jz}%
  \BibitemOpen
  \bibfield  {author} {\bibinfo {author} {\bibfnamefont {A.}~\bibnamefont
  {Dobado}}, \bibinfo {author} {\bibfnamefont {M.~J.}\ \bibnamefont {Herrero}},
  \bibinfo {author} {\bibfnamefont {W.}~\bibnamefont {Hollik}}, \ and\ \bibinfo
  {author} {\bibfnamefont {S.}~\bibnamefont {Penaranda}},\ }\href {\doibase
  10.1103/PhysRevD.66.095016} {\bibfield  {journal} {\bibinfo  {journal} {Phys.
  Rev.}\ }\textbf {\bibinfo {volume} {D66}},\ \bibinfo {pages} {095016}
  (\bibinfo {year} {2002})},\ \Eprint {http://arxiv.org/abs/hep-ph/0208014}
  {arXiv:hep-ph/0208014} \BibitemShut {NoStop}%
\bibitem [{\citenamefont {Aoki}\ \emph {et~al.}(2013)\citenamefont {Aoki},
  \citenamefont {Kanemura}, \citenamefont {Kikuchi},\ and\ \citenamefont
  {Yagyu}}]{Aoki:2012jj}%
  \BibitemOpen
  \bibfield  {author} {\bibinfo {author} {\bibfnamefont {M.}~\bibnamefont
  {Aoki}}, \bibinfo {author} {\bibfnamefont {S.}~\bibnamefont {Kanemura}},
  \bibinfo {author} {\bibfnamefont {M.}~\bibnamefont {Kikuchi}}, \ and\
  \bibinfo {author} {\bibfnamefont {K.}~\bibnamefont {Yagyu}},\ }\href
  {\doibase 10.1103/PhysRevD.87.015012} {\bibfield  {journal} {\bibinfo
  {journal} {Phys. Rev.}\ }\textbf {\bibinfo {volume} {D87}},\ \bibinfo {pages}
  {015012} (\bibinfo {year} {2013})},\ \Eprint {http://arxiv.org/abs/1211.6029}
  {arXiv:1211.6029 [hep-ph]} \BibitemShut {NoStop}%
\bibitem [{\citenamefont {Kanemura}\ \emph {et~al.}(2015)\citenamefont
  {Kanemura}, \citenamefont {Kikuchi},\ and\ \citenamefont
  {Yagyu}}]{Kanemura:2015mxa}%
  \BibitemOpen
  \bibfield  {author} {\bibinfo {author} {\bibfnamefont {S.}~\bibnamefont
  {Kanemura}}, \bibinfo {author} {\bibfnamefont {M.}~\bibnamefont {Kikuchi}}, \
  and\ \bibinfo {author} {\bibfnamefont {K.}~\bibnamefont {Yagyu}},\ }\href
  {\doibase 10.1016/j.nuclphysb.2015.04.015} {\bibfield  {journal} {\bibinfo
  {journal} {Nucl. Phys.}\ }\textbf {\bibinfo {volume} {B896}},\ \bibinfo
  {pages} {80} (\bibinfo {year} {2015})},\ \Eprint
  {http://arxiv.org/abs/1502.07716} {arXiv:1502.07716 [hep-ph]} \BibitemShut
  {NoStop}%
\bibitem [{\citenamefont {Arhrib}\ \emph {et~al.}(2015)\citenamefont {Arhrib},
  \citenamefont {Benbrik}, \citenamefont {El~Falaki},\ and\ \citenamefont
  {Jueid}}]{Arhrib:2015hoa}%
  \BibitemOpen
  \bibfield  {author} {\bibinfo {author} {\bibfnamefont {A.}~\bibnamefont
  {Arhrib}}, \bibinfo {author} {\bibfnamefont {R.}~\bibnamefont {Benbrik}},
  \bibinfo {author} {\bibfnamefont {J.}~\bibnamefont {El~Falaki}}, \ and\
  \bibinfo {author} {\bibfnamefont {A.}~\bibnamefont {Jueid}},\ }\href
  {\doibase 10.1007/JHEP12(2015)007} {\bibfield  {journal} {\bibinfo  {journal}
  {JHEP}\ }\textbf {\bibinfo {volume} {12}},\ \bibinfo {pages} {007} (\bibinfo
  {year} {2015})},\ \Eprint {http://arxiv.org/abs/1507.03630} {arXiv:1507.03630
  [hep-ph]} \BibitemShut {NoStop}%
\bibitem [{\citenamefont {Kanemura}\ \emph
  {et~al.}(2016{\natexlab{a}})\citenamefont {Kanemura}, \citenamefont
  {Kikuchi},\ and\ \citenamefont {Yagyu}}]{Kanemura:2015fra}%
  \BibitemOpen
  \bibfield  {author} {\bibinfo {author} {\bibfnamefont {S.}~\bibnamefont
  {Kanemura}}, \bibinfo {author} {\bibfnamefont {M.}~\bibnamefont {Kikuchi}}, \
  and\ \bibinfo {author} {\bibfnamefont {K.}~\bibnamefont {Yagyu}},\ }\href
  {\doibase 10.1016/j.nuclphysb.2016.04.005} {\bibfield  {journal} {\bibinfo
  {journal} {Nucl. Phys.}\ }\textbf {\bibinfo {volume} {B907}},\ \bibinfo
  {pages} {286} (\bibinfo {year} {2016}{\natexlab{a}})},\ \Eprint
  {http://arxiv.org/abs/1511.06211} {arXiv:1511.06211 [hep-ph]} \BibitemShut
  {NoStop}%
\bibitem [{\citenamefont {Krause}\ \emph {et~al.}(2016)\citenamefont {Krause},
  \citenamefont {Lorenz}, \citenamefont {Mühlleitner}, \citenamefont {Santos},\
  and\ \citenamefont {Ziesche}}]{Krause:2016oke}%
  \BibitemOpen
  \bibfield  {author} {\bibinfo {author} {\bibfnamefont {M.}~\bibnamefont
  {Krause}}, \bibinfo {author} {\bibfnamefont {R.}~\bibnamefont {Lorenz}},
  \bibinfo {author} {\bibfnamefont {M.}~\bibnamefont {Mühlleitner}}, \bibinfo
  {author} {\bibfnamefont {R.}~\bibnamefont {Santos}}, \ and\ \bibinfo {author}
  {\bibfnamefont {H.}~\bibnamefont {Ziesche}},\ }\href {\doibase
  10.1007/JHEP09(2016)143} {\bibfield  {journal} {\bibinfo  {journal} {JHEP}\
  }\textbf {\bibinfo {volume} {09}},\ \bibinfo {pages} {143} (\bibinfo {year}
  {2016})},\ \Eprint {http://arxiv.org/abs/1605.04853} {arXiv:1605.04853
  [hep-ph]} \BibitemShut {NoStop}%
\bibitem [{\citenamefont {Kanemura}\ \emph
  {et~al.}(2016{\natexlab{b}})\citenamefont {Kanemura}, \citenamefont
  {Kikuchi},\ and\ \citenamefont {Sakurai}}]{Kanemura:2016sos}%
  \BibitemOpen
  \bibfield  {author} {\bibinfo {author} {\bibfnamefont {S.}~\bibnamefont
  {Kanemura}}, \bibinfo {author} {\bibfnamefont {M.}~\bibnamefont {Kikuchi}}, \
  and\ \bibinfo {author} {\bibfnamefont {K.}~\bibnamefont {Sakurai}},\ }\href
  {\doibase 10.1103/PhysRevD.94.115011} {\bibfield  {journal} {\bibinfo
  {journal} {Phys. Rev.}\ }\textbf {\bibinfo {volume} {D94}},\ \bibinfo {pages}
  {115011} (\bibinfo {year} {2016}{\natexlab{b}})},\ \Eprint
  {http://arxiv.org/abs/1605.08520} {arXiv:1605.08520 [hep-ph]} \BibitemShut
  {NoStop}%
\bibitem [{\citenamefont {He}\ and\ \citenamefont {Zhu}(2017)}]{He:2016sqr}%
  \BibitemOpen
  \bibfield  {author} {\bibinfo {author} {\bibfnamefont {S.-P.}\ \bibnamefont
  {He}}\ and\ \bibinfo {author} {\bibfnamefont {S.-h.}\ \bibnamefont {Zhu}},\
  }\href {\doibase 10.1016/j.physletb.2016.11.007} {\bibfield  {journal}
  {\bibinfo  {journal} {Phys. Lett.}\ }\textbf {\bibinfo {volume} {B764}},\
  \bibinfo {pages} {31} (\bibinfo {year} {2017})},\ \Eprint
  {http://arxiv.org/abs/1607.04497} {arXiv:1607.04497 [hep-ph]} \BibitemShut
  {NoStop}%
\bibitem [{\citenamefont {Kanemura}\ \emph
  {et~al.}(2017{\natexlab{a}})\citenamefont {Kanemura}, \citenamefont
  {Kikuchi},\ and\ \citenamefont {Yagyu}}]{Kanemura:2016lkz}%
  \BibitemOpen
  \bibfield  {author} {\bibinfo {author} {\bibfnamefont {S.}~\bibnamefont
  {Kanemura}}, \bibinfo {author} {\bibfnamefont {M.}~\bibnamefont {Kikuchi}}, \
  and\ \bibinfo {author} {\bibfnamefont {K.}~\bibnamefont {Yagyu}},\ }\href
  {\doibase 10.1016/j.nuclphysb.2017.02.004} {\bibfield  {journal} {\bibinfo
  {journal} {Nucl. Phys.}\ }\textbf {\bibinfo {volume} {B917}},\ \bibinfo
  {pages} {154} (\bibinfo {year} {2017}{\natexlab{a}})},\ \Eprint
  {http://arxiv.org/abs/1608.01582} {arXiv:1608.01582 [hep-ph]} \BibitemShut
  {NoStop}%
\bibitem [{\citenamefont {Kanemura}\ \emph
  {et~al.}(2017{\natexlab{b}})\citenamefont {Kanemura}, \citenamefont
  {Kikuchi}, \citenamefont {Sakurai},\ and\ \citenamefont
  {Yagyu}}]{Kanemura:2017wtm}%
  \BibitemOpen
  \bibfield  {author} {\bibinfo {author} {\bibfnamefont {S.}~\bibnamefont
  {Kanemura}}, \bibinfo {author} {\bibfnamefont {M.}~\bibnamefont {Kikuchi}},
  \bibinfo {author} {\bibfnamefont {K.}~\bibnamefont {Sakurai}}, \ and\
  \bibinfo {author} {\bibfnamefont {K.}~\bibnamefont {Yagyu}},\ }\href
  {\doibase 10.1103/PhysRevD.96.035014} {\bibfield  {journal} {\bibinfo
  {journal} {Phys. Rev.}\ }\textbf {\bibinfo {volume} {D96}},\ \bibinfo {pages}
  {035014} (\bibinfo {year} {2017}{\natexlab{b}})},\ \Eprint
  {http://arxiv.org/abs/1705.05399} {arXiv:1705.05399 [hep-ph]} \BibitemShut
  {NoStop}%
\bibitem [{\citenamefont {Kanemura}\ \emph {et~al.}(2018)\citenamefont
  {Kanemura}, \citenamefont {Kikuchi}, \citenamefont {Sakurai},\ and\
  \citenamefont {Yagyu}}]{Kanemura:2017gbi}%
  \BibitemOpen
  \bibfield  {author} {\bibinfo {author} {\bibfnamefont {S.}~\bibnamefont
  {Kanemura}}, \bibinfo {author} {\bibfnamefont {M.}~\bibnamefont {Kikuchi}},
  \bibinfo {author} {\bibfnamefont {K.}~\bibnamefont {Sakurai}}, \ and\
  \bibinfo {author} {\bibfnamefont {K.}~\bibnamefont {Yagyu}},\ }\href
  {\doibase 10.1016/j.cpc.2018.06.012} {\bibfield  {journal} {\bibinfo
  {journal} {Comput. Phys. Commun.}\ }\textbf {\bibinfo {volume} {233}},\
  \bibinfo {pages} {134} (\bibinfo {year} {2018})},\ \Eprint
  {http://arxiv.org/abs/1710.04603} {arXiv:1710.04603 [hep-ph]} \BibitemShut
  {NoStop}%
\bibitem [{\citenamefont {Kanemura}\ \emph
  {et~al.}(2019{\natexlab{a}})\citenamefont {Kanemura}, \citenamefont
  {Kikuchi}, \citenamefont {Mawatari}, \citenamefont {Sakurai},\ and\
  \citenamefont {Yagyu}}]{Kanemura:2019slf}%
  \BibitemOpen
  \bibfield  {author} {\bibinfo {author} {\bibfnamefont {S.}~\bibnamefont
  {Kanemura}}, \bibinfo {author} {\bibfnamefont {M.}~\bibnamefont {Kikuchi}},
  \bibinfo {author} {\bibfnamefont {K.}~\bibnamefont {Mawatari}}, \bibinfo
  {author} {\bibfnamefont {K.}~\bibnamefont {Sakurai}}, \ and\ \bibinfo
  {author} {\bibfnamefont {K.}~\bibnamefont {Yagyu}},\ }\href@noop {} {\
  (\bibinfo {year} {2019}{\natexlab{a}})},\ \Eprint
  {http://arxiv.org/abs/1910.12769} {arXiv:1910.12769 [hep-ph]} \BibitemShut
  {NoStop}%
\bibitem [{\citenamefont {Krause}\ \emph {et~al.}(2018)\citenamefont {Krause},
  \citenamefont {Mühlleitner},\ and\ \citenamefont {Spira}}]{Krause:2018wmo}%
  \BibitemOpen
  \bibfield  {author} {\bibinfo {author} {\bibfnamefont {M.}~\bibnamefont
  {Krause}}, \bibinfo {author} {\bibfnamefont {M.}~\bibnamefont
  {Mühlleitner}}, \ and\ \bibinfo {author} {\bibfnamefont {M.}~\bibnamefont
  {Spira}},\ }\href {\doibase 10.1016/j.cpc.2019.08.003} {\  (\bibinfo {year}
  {2018}),\ 10.1016/j.cpc.2019.08.003},\ \Eprint
  {http://arxiv.org/abs/1810.00768} {arXiv:1810.00768 [hep-ph]} \BibitemShut
  {NoStop}%
\bibitem [{\citenamefont {Brucherseifer}\ \emph {et~al.}(2014)\citenamefont
  {Brucherseifer}, \citenamefont {Gavin},\ and\ \citenamefont
  {Spira}}]{Brucherseifer:2013qva}%
  \BibitemOpen
  \bibfield  {author} {\bibinfo {author} {\bibfnamefont {M.}~\bibnamefont
  {Brucherseifer}}, \bibinfo {author} {\bibfnamefont {R.}~\bibnamefont
  {Gavin}}, \ and\ \bibinfo {author} {\bibfnamefont {M.}~\bibnamefont
  {Spira}},\ }\href {\doibase 10.1103/PhysRevD.90.117701} {\bibfield  {journal}
  {\bibinfo  {journal} {Phys. Rev.}\ }\textbf {\bibinfo {volume} {D90}},\
  \bibinfo {pages} {117701} (\bibinfo {year} {2014})},\ \Eprint
  {http://arxiv.org/abs/1309.3140} {arXiv:1309.3140 [hep-ph]} \BibitemShut
  {NoStop}%
\bibitem [{\citenamefont {Mühlleitner}\ \emph {et~al.}(2015)\citenamefont
  {Mühlleitner}, \citenamefont {Nhung},\ and\ \citenamefont
  {Ziesche}}]{Muhlleitner:2015dua}%
  \BibitemOpen
  \bibfield  {author} {\bibinfo {author} {\bibfnamefont {M.}~\bibnamefont
  {Mühlleitner}}, \bibinfo {author} {\bibfnamefont {D.~T.}\ \bibnamefont
  {Nhung}}, \ and\ \bibinfo {author} {\bibfnamefont {H.}~\bibnamefont
  {Ziesche}},\ }\href {\doibase 10.1007/JHEP12(2015)034} {\bibfield  {journal}
  {\bibinfo  {journal} {JHEP}\ }\textbf {\bibinfo {volume} {12}},\ \bibinfo
  {pages} {034} (\bibinfo {year} {2015})},\ \Eprint
  {http://arxiv.org/abs/1506.03321} {arXiv:1506.03321 [hep-ph]} \BibitemShut
  {NoStop}%
\bibitem [{\citenamefont {Senaha}(2019)}]{Senaha:2018xek}%
  \BibitemOpen
  \bibfield  {author} {\bibinfo {author} {\bibfnamefont {E.}~\bibnamefont
  {Senaha}},\ }\href {\doibase 10.1103/PhysRevD.100.055034} {\bibfield
  {journal} {\bibinfo  {journal} {Phys. Rev.}\ }\textbf {\bibinfo {volume}
  {D100}},\ \bibinfo {pages} {055034} (\bibinfo {year} {2019})},\ \Eprint
  {http://arxiv.org/abs/1811.00336} {arXiv:1811.00336 [hep-ph]} \BibitemShut
  {NoStop}%
\bibitem [{\citenamefont {Basler}\ \emph {et~al.}(2017)\citenamefont {Basler},
  \citenamefont {Krause}, \citenamefont {Mühlleitner}, \citenamefont
  {Wittbrodt},\ and\ \citenamefont {Wlotzka}}]{Basler:2016obg}%
  \BibitemOpen
  \bibfield  {author} {\bibinfo {author} {\bibfnamefont {P.}~\bibnamefont
  {Basler}}, \bibinfo {author} {\bibfnamefont {M.}~\bibnamefont {Krause}},
  \bibinfo {author} {\bibfnamefont {M.}~\bibnamefont {Mühlleitner}}, \bibinfo
  {author} {\bibfnamefont {J.}~\bibnamefont {Wittbrodt}}, \ and\ \bibinfo
  {author} {\bibfnamefont {A.}~\bibnamefont {Wlotzka}},\ }\href {\doibase
  10.1007/JHEP02(2017)121} {\bibfield  {journal} {\bibinfo  {journal} {JHEP}\
  }\textbf {\bibinfo {volume} {02}},\ \bibinfo {pages} {121} (\bibinfo {year}
  {2017})},\ \Eprint {http://arxiv.org/abs/1612.04086} {arXiv:1612.04086
  [hep-ph]} \BibitemShut {NoStop}%
\bibitem [{\citenamefont {Braathen}\ \emph {et~al.}(2017)\citenamefont
  {Braathen}, \citenamefont {Goodsell},\ and\ \citenamefont
  {Staub}}]{Braathen:2017izn}%
  \BibitemOpen
  \bibfield  {author} {\bibinfo {author} {\bibfnamefont {J.}~\bibnamefont
  {Braathen}}, \bibinfo {author} {\bibfnamefont {M.~D.}\ \bibnamefont
  {Goodsell}}, \ and\ \bibinfo {author} {\bibfnamefont {F.}~\bibnamefont
  {Staub}},\ }\href {\doibase 10.1140/epjc/s10052-017-5303-x} {\bibfield
  {journal} {\bibinfo  {journal} {Eur. Phys. J.}\ }\textbf {\bibinfo {volume}
  {C77}},\ \bibinfo {pages} {757} (\bibinfo {year} {2017})},\ \Eprint
  {http://arxiv.org/abs/1706.05372} {arXiv:1706.05372 [hep-ph]} \BibitemShut
  {NoStop}%
\bibitem [{\citenamefont {Braathen}\ \emph {et~al.}(2018)\citenamefont
  {Braathen}, \citenamefont {Goodsell}, \citenamefont {Krauss}, \citenamefont
  {Opferkuch},\ and\ \citenamefont {Staub}}]{Braathen:2017jvs}%
  \BibitemOpen
  \bibfield  {author} {\bibinfo {author} {\bibfnamefont {J.}~\bibnamefont
  {Braathen}}, \bibinfo {author} {\bibfnamefont {M.~D.}\ \bibnamefont
  {Goodsell}}, \bibinfo {author} {\bibfnamefont {M.~E.}\ \bibnamefont
  {Krauss}}, \bibinfo {author} {\bibfnamefont {T.}~\bibnamefont {Opferkuch}}, \
  and\ \bibinfo {author} {\bibfnamefont {F.}~\bibnamefont {Staub}},\ }\href
  {\doibase 10.1103/PhysRevD.97.015011} {\bibfield  {journal} {\bibinfo
  {journal} {Phys. Rev.}\ }\textbf {\bibinfo {volume} {D97}},\ \bibinfo {pages}
  {015011} (\bibinfo {year} {2018})},\ \Eprint
  {http://arxiv.org/abs/1711.08460} {arXiv:1711.08460 [hep-ph]} \BibitemShut
  {NoStop}%
\bibitem [{\citenamefont {Basler}\ \emph {et~al.}(2018)\citenamefont {Basler},
  \citenamefont {Mühlleitner},\ and\ \citenamefont
  {Wittbrodt}}]{Basler:2017uxn}%
  \BibitemOpen
  \bibfield  {author} {\bibinfo {author} {\bibfnamefont {P.}~\bibnamefont
  {Basler}}, \bibinfo {author} {\bibfnamefont {M.}~\bibnamefont
  {Mühlleitner}}, \ and\ \bibinfo {author} {\bibfnamefont {J.}~\bibnamefont
  {Wittbrodt}},\ }\href {\doibase 10.1007/JHEP03(2018)061} {\bibfield
  {journal} {\bibinfo  {journal} {JHEP}\ }\textbf {\bibinfo {volume} {03}},\
  \bibinfo {pages} {061} (\bibinfo {year} {2018})},\ \Eprint
  {http://arxiv.org/abs/1711.04097} {arXiv:1711.04097 [hep-ph]} \BibitemShut
  {NoStop}%
\bibitem [{\citenamefont {Kanemura}\ \emph
  {et~al.}(2019{\natexlab{b}})\citenamefont {Kanemura}, \citenamefont
  {Kikuchi}, \citenamefont {Mawatari}, \citenamefont {Sakurai},\ and\
  \citenamefont {Yagyu}}]{Kanemura:2019kjg}%
  \BibitemOpen
  \bibfield  {author} {\bibinfo {author} {\bibfnamefont {S.}~\bibnamefont
  {Kanemura}}, \bibinfo {author} {\bibfnamefont {M.}~\bibnamefont {Kikuchi}},
  \bibinfo {author} {\bibfnamefont {K.}~\bibnamefont {Mawatari}}, \bibinfo
  {author} {\bibfnamefont {K.}~\bibnamefont {Sakurai}}, \ and\ \bibinfo
  {author} {\bibfnamefont {K.}~\bibnamefont {Yagyu}},\ }\href {\doibase
  10.1016/j.nuclphysb.2019.114791} {\bibfield  {journal} {\bibinfo  {journal}
  {Nucl. Phys.}\ }\textbf {\bibinfo {volume} {B949}},\ \bibinfo {pages}
  {114791} (\bibinfo {year} {2019}{\natexlab{b}})},\ \Eprint
  {http://arxiv.org/abs/1906.10070} {arXiv:1906.10070 [hep-ph]} \BibitemShut
  {NoStop}%
\bibitem [{\citenamefont {Lee}\ \emph {et~al.}(1977)\citenamefont {Lee},
  \citenamefont {Quigg},\ and\ \citenamefont {Thacker}}]{Lee:1977eg}%
  \BibitemOpen
  \bibfield  {author} {\bibinfo {author} {\bibfnamefont {B.~W.}\ \bibnamefont
  {Lee}}, \bibinfo {author} {\bibfnamefont {C.}~\bibnamefont {Quigg}}, \ and\
  \bibinfo {author} {\bibfnamefont {H.~B.}\ \bibnamefont {Thacker}},\ }\href
  {\doibase 10.1103/PhysRevD.16.1519} {\bibfield  {journal} {\bibinfo
  {journal} {Phys. Rev.}\ }\textbf {\bibinfo {volume} {D16}},\ \bibinfo {pages}
  {1519} (\bibinfo {year} {1977})}\BibitemShut {NoStop}%
\bibitem [{\citenamefont {Kanemura}\ \emph {et~al.}(1993)\citenamefont
  {Kanemura}, \citenamefont {Kubota},\ and\ \citenamefont
  {Takasugi}}]{Kanemura:1993hm}%
  \BibitemOpen
  \bibfield  {author} {\bibinfo {author} {\bibfnamefont {S.}~\bibnamefont
  {Kanemura}}, \bibinfo {author} {\bibfnamefont {T.}~\bibnamefont {Kubota}}, \
  and\ \bibinfo {author} {\bibfnamefont {E.}~\bibnamefont {Takasugi}},\ }\href
  {\doibase 10.1016/0370-2693(93)91205-2} {\bibfield  {journal} {\bibinfo
  {journal} {Phys. Lett.}\ }\textbf {\bibinfo {volume} {B313}},\ \bibinfo
  {pages} {155} (\bibinfo {year} {1993})},\ \Eprint
  {http://arxiv.org/abs/hep-ph/9303263} {arXiv:hep-ph/9303263} \BibitemShut
  {NoStop}%
\bibitem [{\citenamefont {Akeroyd}\ \emph {et~al.}(2000)\citenamefont
  {Akeroyd}, \citenamefont {Arhrib},\ and\ \citenamefont
  {Naimi}}]{Akeroyd:2000wc}%
  \BibitemOpen
  \bibfield  {author} {\bibinfo {author} {\bibfnamefont {A.~G.}\ \bibnamefont
  {Akeroyd}}, \bibinfo {author} {\bibfnamefont {A.}~\bibnamefont {Arhrib}}, \
  and\ \bibinfo {author} {\bibfnamefont {E.-M.}\ \bibnamefont {Naimi}},\ }\href
  {\doibase 10.1016/S0370-2693(00)00962-X} {\bibfield  {journal} {\bibinfo
  {journal} {Phys. Lett.}\ }\textbf {\bibinfo {volume} {B490}},\ \bibinfo
  {pages} {119} (\bibinfo {year} {2000})},\ \Eprint
  {http://arxiv.org/abs/hep-ph/0006035} {arXiv:hep-ph/0006035 [hep-ph]}
  \BibitemShut {NoStop}%
\bibitem [{\citenamefont {Martin}(2002)}]{Martin:2001vx}%
  \BibitemOpen
  \bibfield  {author} {\bibinfo {author} {\bibfnamefont {S.~P.}\ \bibnamefont
  {Martin}},\ }\href {\doibase 10.1103/PhysRevD.65.116003} {\bibfield
  {journal} {\bibinfo  {journal} {Phys. Rev.}\ }\textbf {\bibinfo {volume}
  {D65}},\ \bibinfo {pages} {116003} (\bibinfo {year} {2002})},\ \Eprint
  {http://arxiv.org/abs/hep-ph/0111209} {arXiv:hep-ph/0111209} \BibitemShut
  {NoStop}%
\bibitem [{\citenamefont {Martin}(2003)}]{Martin:2003qz}%
  \BibitemOpen
  \bibfield  {author} {\bibinfo {author} {\bibfnamefont {S.~P.}\ \bibnamefont
  {Martin}},\ }\href {\doibase 10.1103/PhysRevD.68.075002} {\bibfield
  {journal} {\bibinfo  {journal} {Phys. Rev.}\ }\textbf {\bibinfo {volume}
  {D68}},\ \bibinfo {pages} {075002} (\bibinfo {year} {2003})},\ \Eprint
  {http://arxiv.org/abs/hep-ph/0307101} {arXiv:hep-ph/0307101} \BibitemShut
  {NoStop}%
\bibitem [{\citenamefont {Degrassi}\ and\ \citenamefont
  {Slavich}(2010)}]{Degrassi:2009yq}%
  \BibitemOpen
  \bibfield  {author} {\bibinfo {author} {\bibfnamefont {G.}~\bibnamefont
  {Degrassi}}\ and\ \bibinfo {author} {\bibfnamefont {P.}~\bibnamefont
  {Slavich}},\ }\href {\doibase 10.1016/j.nuclphysb.2009.09.018} {\bibfield
  {journal} {\bibinfo  {journal} {Nucl. Phys.}\ }\textbf {\bibinfo {volume}
  {B825}},\ \bibinfo {pages} {119} (\bibinfo {year} {2010})},\ \Eprint
  {http://arxiv.org/abs/0907.4682} {arXiv:0907.4682 [hep-ph]} \BibitemShut
  {NoStop}%
\bibitem [{\citenamefont {Braathen}\ and\ \citenamefont
  {Goodsell}(2016)}]{Braathen:2016cqe}%
  \BibitemOpen
  \bibfield  {author} {\bibinfo {author} {\bibfnamefont {J.}~\bibnamefont
  {Braathen}}\ and\ \bibinfo {author} {\bibfnamefont {M.~D.}\ \bibnamefont
  {Goodsell}},\ }\href {\doibase 10.1007/JHEP12(2016)056} {\bibfield  {journal}
  {\bibinfo  {journal} {JHEP}\ }\textbf {\bibinfo {volume} {12}},\ \bibinfo
  {pages} {056} (\bibinfo {year} {2016})},\ \Eprint
  {http://arxiv.org/abs/1609.06977} {arXiv:1609.06977 [hep-ph]} \BibitemShut
  {NoStop}%
\end{thebibliography}%

\end{document}